# Backward propagating quantum repeater protocol with multiple quantum memories

Yuhei Sekiguchi,[1]* Satsuki Okumura,[2,3] and Hideo Kosaka[1,3][†]

[1]Institute of Advanced Sciences, Yokohama National University, 79-5 Tokiwadai, Hodogaya, Yokohama 240-8501, Japan

[2]Department of Advanced Energy, Graduate School of Frontier Sciences, University of Tokyo, 5-1-5 Kashiwanoha, Kashiwa, Chiba, 277-8561, Japan

[3]Department of Physics, Graduate School of Engineering Science, Yokohama National University, 79-5 Tokiwadai, Hodogaya, Yokohama 240-8501, Japan

*yuhei-sekiguchi-ct@ynu.ac.jp

[†]kosaka-hideo-yp@ynu.ac.jp

**Abstract**

Quantum repeaters with multiple quantum memories provide high throughput, low latency, and high fidelity quantum state (qubit) transfer over long distances. However, conventional quantum repeater protocols require full connections among the multiple quantum memories in a repeater node, which is technically challenging. Here, we propose a quantum repeater protocol based on backward propagating photon emission and absorption, where the quantum memories are multiplexed in the time-domain to speed up a single transmission channel without requiring full connectivity, drastically facilitating physical implementation. Although the protocol is described with nitrogen-vacancy (NV) centers in diamond, it is applicable to various physical systems and opens up the possibility of high-speed high-fidelity quantum networks for distributed quantum computation and quantum Internet.

**Introduction**

Quantum communication has the potential to enable not only quantum cryptography for the secure exchange of classical information [1], but also distributed quantum computers that scale up the quantum system to accelerate the quantum computation [2]. Direct quantum communication succeeds only when a photon is transmitted without loss over optical fiber from a sender to a receiver, while the success probability decreases exponentially with a distance of $L$, making it impractical for long-distance communication. This photon loss can be overcome by using quantum repeaters, which generate probabilistic entanglement via photons between quantum memories in adjacent nodes and deterministically teleport a qubit to extend the entanglement. By repeating this process, we can establish long-distance entanglement, which is used as a resource for quantum key distribution, blind quantum computation, and distributed quantum computation [3–5].

The current quantum repeater protocols [6–17] are technically challenging or unable to achieve high throughput rates. For example, photon loss not only in the fiber but also in various components, such as optical coupling between a quantum memory and a transmission channel [18–31] and quantum frequency conversion between the memory frequency and the communication frequency [32–35], reduce the success probability of entanglement in adjacent nodes to suppress the throughput. Entangling attempts are thus repeated many times until success, even though the repetition time is limited by the photon round-trip time including the return time of the heralding signal upon success. One approach to solve this problem is simply parallelizing multiple channels to increase the throughput. However, since the time for transferring the qubit over a long distance in the individual channel remains long, the quantum memories, where the qubit passes through, require a similarly long memory time. Another approach utilizing multiple memories has been proposed to solve this problem by dynamically connecting arbitrary two memories that have succeeded to entangle both adjacent nodes [10–12] and by encoding multiple memories or photons into fault-tolerant states [13–17], while it is hard to reasonably implement due to the requirement of full and on-demand connectivity in the node or near-unity efficiency of the optical interface. For example, the proposed protocol using weakly coupled nuclear spin memories with an NV center does not offer dramatic gains because the operation time is too long [8].

In this paper, we propose a backward propagating quantum repeater protocol that allows multiple memories to concentrate on a single transmission channel by simply routing photons emitted from the memories into one previous node. The protocol drastically reduces the required memory time and eliminates the need for full connectivity, resulting in high throughput and high fidelity with simple implementation. Although the protocol can be implemented in various quantum systems, including atoms, ions, quantum dots, and diamond color centers, this paper specifically describes how to use an electron in an NV center in diamond as the interface between photon and quantum memories consisting of nuclear spins near the NV center.

**Parallelized Memory Protocol**

We first describe a backward propagating parallelized memory protocol to characterize a backward propagating multiplexed memory protocol described later. Independent channels, which are parallelized into $M$ channels, are connected by $N$ NVs in series from a sender (Alice) to a receiver (Bob). The distance between adjacent nodes is $L_0$, and an NV has two quantum memories. We assume that the quantum teleportation (repeating) process takes from left to right as shown in Fig. 1(a) for arbitrary adjacent nodes. Entanglement between a memory and a photon is first generated in the next (right) node. The photon is then transferred to the memory in the previous (left) node through an optical fiber, resulting in the entanglement between the adjacent nodes. This entanglement is used to transfer a qubit to the next node by quantum teleportation with Bell state measurement. By repeating the teleportation process from left to right, the qubit finally arrives at Bob [Fig 1(b)]. If the qubit sent by Alice is one side of an entangled pair, each quantum teleportation process corresponds to entanglement swapping. Provided that the probabilistic entanglement between adjacent nodes is heralded and the Bell state measurement is complete, the qubit transfer is performed deterministically to overcome the transmission losses. In this protocol, the throughput increases in proportion to the number of parallelized channels $M$, while the end-to-end transfer time of the qubit remains unchanged.

**Multiplexed Memory Protocol**

To transfer the qubit over a long distance with high fidelity, the memory time needs to be longer than the end-to-end transfer time, including the time spent waiting at the repeater nodes. We therefore propose the backward propagating multiplexed memory protocol to overcome this problem [Fig. 1(c)]. The significances of the protocol are two folds. First, a photon is transmitted in the opposite way of quantum teleportation, backing toward the previous (left) node. Second, all of the NVs at the next node transmit photons entangled with their memories sequentially with time-domain multiplexing, while the NV at the previous node is restricted to only one holding the qubit to be transferred, where the time-multiplexed photons are routed with optical elements within the node. In this way, the multiple memories are concentrated on the acceleration of a single channel transmission rate. If the time for routing is negligibly short compared to the photon round-trip time, the throughput increases roughly in proportion to the number of NVs in a node, $M$. In addition, the memory time requirement is drastically relaxed since the time required for quantum teleportation to the adjacent node is reduced, and thus higher fidelity is achieved. Note that this protocol cannot be realized by forward propagation of photons.

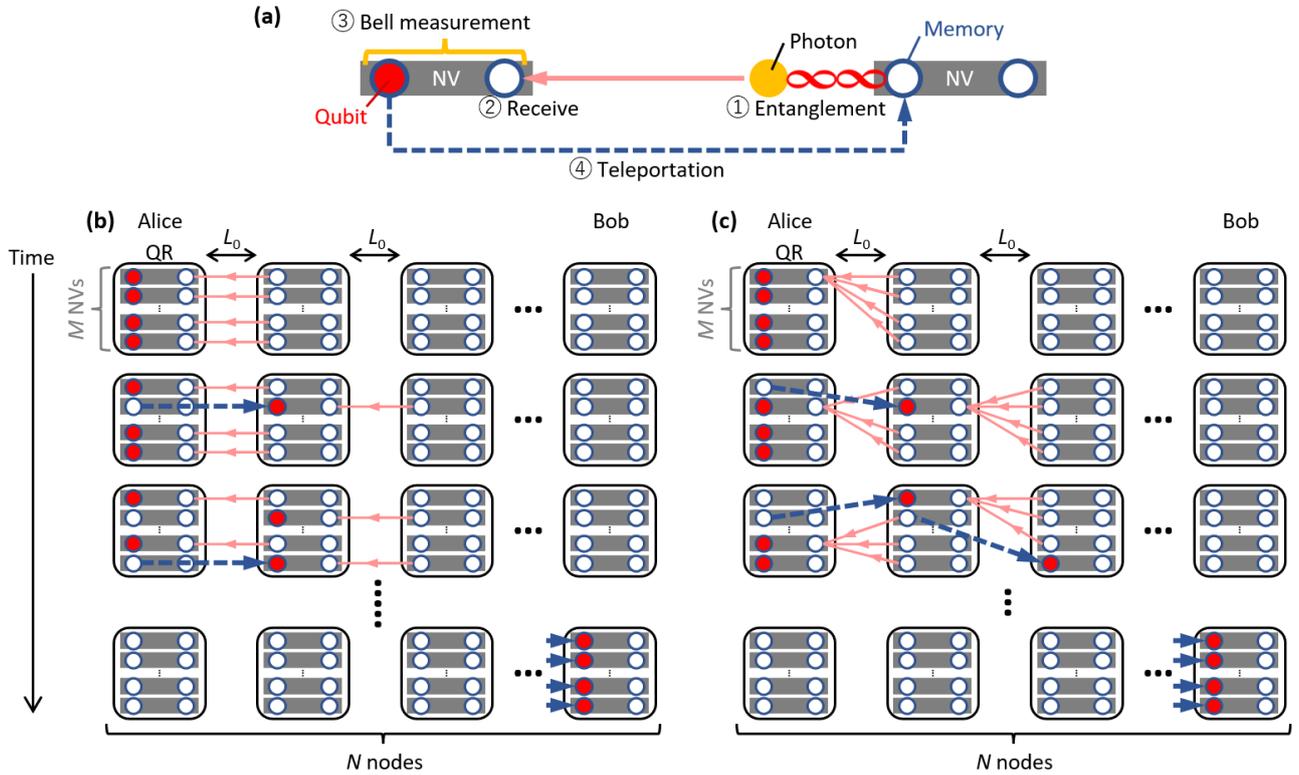

**Fig. 1. Overview of backward propagating quantum repeater protocols. (a) A process for quantum teleporting (repeating) a qubit (red circle) between two nitrogen-vacancy (NV) centers (gray rectangles). An NV has two quantum memories (blue circle frames). Details are written in the main text. (b) Schematic time evolution of parallelized memory protocol. (c) Schematic time evolution of multiplexed memory protocol. $N$ denotes the number of quantum repeater (QR) nodes, $M$ denotes the number of NVs implemented at each node, and $L_0$ denotes the distance between the adjacent nodes. The leftmost is the sender (Alice) and the rightmost is the receiver (Bob). The photon transmission is indicated by the pink solid arrows, and the qubit being transferred is indicated by the blue dashed arrows.**

**Numerical analysis**

The performance of our protocol is determined by the number of nodes $N$, the number of NVs in a node $M$, the success probability of entangling adjacent nodes including transmission loss $P$, and the repetition period for entangling adjacent nodes including the heralding time with neglecting the routing time $t_0$. We here assume that the NV holding a qubit for being transferred cannot emit the entangled photon. The difficulty of modeling quantum repeater protocols in time sequence arises from the probabilistic evolution of the network, where the heralding time changes in every event. Since it is necessary to organize the traffic in the presence of communication congestion, it is very difficult to gain algebraic insight.

Numerical simulation using the Monte Carlo (MC) method is an effective way to accurately analyze such networks. Figure 2 shows the results of the MC simulations of the parallelized and multiplexed memory protocols. Note that the simulation ignores the fidelity degradation of qubits during the entire process. Figures 2(a) and 2(b) show the number of completed end-to-end transfers as a function of total time, which indicates two significant differences between the two protocols. First, the latency defined by the waiting time until the first qubit arrives at Bob is approximately $M$-times shorter in the multiplexed memory protocol than that in the parallelized memory protocol as shown in Fig. 2(a). Second, the throughput defined by the number of completed end-to-end qubit transfers per second in the steady-state is approximately doubled in the multiplexed memory protocol by that in the parallelized memory protocol as shown in Fig. 2(b). The throughput is proportional to not only the number of entangling attempts but also the number of NVs that can emit photons. In the parallelized memory protocol, the number of NVs that can emit photons is equal to the number of NVs that can receive photons, which asymptotically approaches half of the total memory number $NM/2$. On the other hand, in the multiplexed memory protocol, the number of NVs that can emit photons asymptotically approaches the total memory number $NM$, resulting in twice the throughput of the parallelized memory protocol. Furthermore, the shorter latency and higher throughput significantly relax the requirement for long memory time. As shown in Fig. 2(c), the average time for end-to-end transfer and the memory time required to store a qubit is almost inversely proportional to the number of multiplexed memories, although $M$-fold improvement is not ideally achieved due to traffic congestion of qubit.

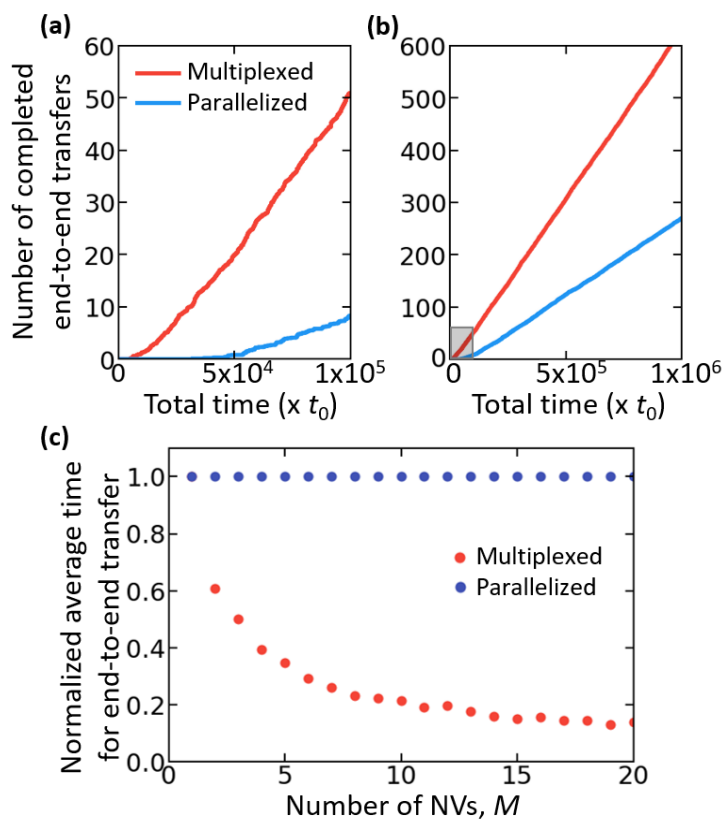

**Fig. 2. Numerical simulation of the protocols. (a, b)** The number of completed end-to-end transfers as a function of total time, showing in short (a) and long term (b) ranges. The gray shaded region in (b) indicates the range in (a). We assume the number of nodes $N=10$, the number of NVs at a node $M=10$, and the probability of entangling adjacent nodes $P=0.0001$. **(c)** The normalized average time for end-to-end transfer of each qubit. We assume $N=10$ and $P=0.0001$. Blue lines (dots) show parallelized memory protocol and red lines (dots) for multiplexed memory protocol.

**Experimental Implementation**

For experimental implementation, we propose an emission-absorption-based scheme for heralded entanglement generation between adjacent nodes [Fig 3a] using two NVs. Triplet states of the negatively charged NV center have a $\Lambda$-shaped three-level structure. One of the excited states, $|A_2\rangle$, is an eigenstate with entangled spin and orbital angular momentum corresponding to one of four Bell states, causing correlations between the electron spin $|m_S = \pm 1\rangle_S$ and the photon polarizations $|\pm 1\rangle_P$ upon the optical transition [36], where $|+1\rangle_P$ and $|-1\rangle_P$ denote right- and left-circular polarizations. Thus, spontaneous emission from $|A_2\rangle$ generates entanglement between the spin of the remaining electron and the polarization of emitted photon [18,25] [Fig 3b]. If the emitted photon from the next node is absorbed by $|A_2\rangle$ at the previous node, the electron spin and the photon polarization must be in the Bell state in the opposite process of emission. In other words, the detection of a photon absorption implies projection into the Bell state [20]. By preparing the entanglement of the electron and nuclear spins before absorption at the previous node, the photon state can be transferred to the nuclear spin memory in the principle of quantum teleportation [21,23]. If the efficiency of entangled emission and transmission is sufficiently small, the failure of absorption will not affect the prepared entanglement. Therefore, the entanglement can be reused without further initialization while the probability of being absorbed twice is negligible. In practice, the fidelity can be increased by reducing the number of initializations [37,38], since initialization probabilistically dephases the quantum memory waiting for teleportation to the next node. Although the above benefits cannot be obtained, the backward propagating protocol is applicable to the popular intra-node entanglement generation method [19,22,26–29], which emits a photon on both sides and measures them at the intermediate beamsplitter, by moving the beamsplitter position to the photon receiving node, for eliminating the requirement for the full connectivity.

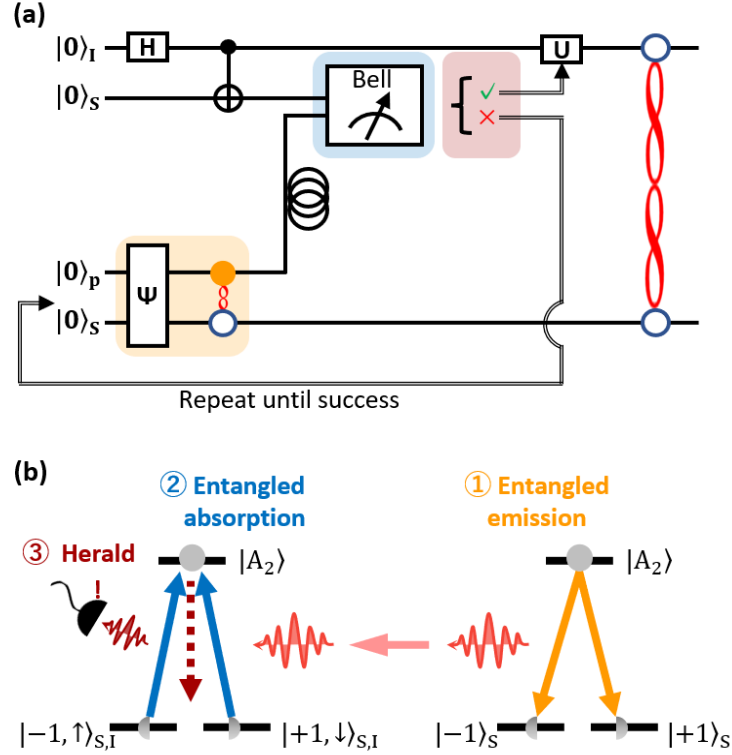

**Fig. 3. Process for heralded entanglement generation between adjacent nodes based on emission and absorption. (a) Quantum circuit diagram. $\Psi$ denotes entanglement generation of $|\Psi\rangle_{S,p} = (|+1,-1\rangle_{S,p} + |-1,+1\rangle_{S,p})/\sqrt{2}$ between an electron spin and emitted photon at the next node. Bell denotes the Bell-state measurement between the electron spin and the absorbed photon at the previous node. H and U respectively denote Hadamard and unitary operations. (b) Schematic energy level diagrams for entangled emission at the next node and entangled absorption at the previous node with heralded function. In the previous node, an entangled state of electron and nuclear spins, $(|+1,\uparrow\rangle_{S,I} + |-1,\downarrow\rangle_{S,I})/\sqrt{2}$, is prepared before the absorption, where $|\uparrow\rangle_I$ and $|\downarrow\rangle_I$ denotes nuclear spin bases.**

**Summary**

In conclusion, we have proposed a novel backward propagating quantum repeater protocol that utilizes multiple quantum memories based on photon emission and absorption to speed up a single transmission channel. The protocol has the potential to significantly increase the entanglement extension speed without fully connecting quantum memories in individual nodes, greatly increasing its practical feasibility and paving the way for high-speed high-fidelity quantum repeater networks with various physical systems.

**Acknowledgments**

We thank Yuichiro Matsuzaki and Kae Nemoto for their discussions. This work was supported by Japan Society for the Promotion of Science (JSPS) Grants-in-Aid for Scientific Research (20H05661, 20K2044120); by Japan Science and Technology Agency (JST) CREST (JPMJCR1773); and by JST Moonshot R&D (JPMJMS2062). We also acknowledge the assistance of the Ministry of Internal Affairs and Communications (MIC) under the initiative Research and Development for Construction of a Global Quantum Cryptography Network (JPMI00316).

**References**

[1] N. Gisin, G. Ribordy, W. Tittel, and H. Zbinden, *Quantum Cryptography*, Reviews of Modern Physics **74**, 145 (2002).

[2] D. Cuomo, M. Caleffi, and A. S. Cacciapuoti, *Towards a Distributed Quantum Computing Ecosystem*, IET Quantum Communication **1**, 3 (2020).

[3] H. J. Kimble, *The Quantum Internet*, Nature **453**, 1023 (2008).

[4] S. Wehner, D. Elkouss, and R. Hanson, *Quantum Internet: A Vision for the Road Ahead*, Science **362**, 6412 (2018).

[5] D. Awschalom et al., *Development of Quantum Interconnects (QuICs) for Next-Generation Information Technologies*, PRX Quantum **2**, 017002 (2021).

[6] L. Childress, J. M. Taylor, A. S. Sørensen, and M. D. Lukin, *Fault-Tolerant Quantum Communication Based on Solid-State Photon Emitters*, Physical Review Letters **96**, 070504 (2006).

[7] B. Scharfenberger, H. Kosaka, W. J. Munro, and K. Nemoto, *Absorption-Based Quantum Communication with NV Centres*, New Journal of Physics **17**, 103012 (2015).

[8] S. B. van Dam, P. C. Humphreys, F. Rozpędek, S. Wehner, and R. Hanson, *Multiplexed Entanglement Generation over Quantum Networks Using Multi-Qubit Nodes*, Quantum Science and Technology **2**, 034002 (2017).

[9] F. Rozpędek, R. Yehia, K. Goodenough, M. Ruf, P. C. Humphreys, R. Hanson, S. Wehner, and D. Elkouss, *Near-Term Quantum-Repeater Experiments with Nitrogen-Vacancy Centers: Overcoming the Limitations of Direct Transmission*, Physical Review A **99**, 052330 (2019).

[10] O. A. Collins, S. D. Jenkins, A. Kuzmich, and T. A. B. Kennedy, *Multiplexed Memory-Insensitive Quantum Repeaters*, Physical Review Letters **98**, 060502 (2007).

[11] M. Razavi, M. Piani, and N. Lütkenhaus, *Quantum Repeaters with Imperfect Memories: Cost and Scalability*, Physical Review A **80**, 032301 (2009).

[12] W. J. Munro, K. A. Harrison, A. M. Stephens, S. J. Devitt, and K. Nemoto, *From Quantum Multiplexing to High-Performance Quantum Networking*, Nature Photonics **4**, 792 (2010).

[13] L. Jiang, J. M. Taylor, K. Nemoto, W. J. Munro, R. van Meter, and M. D. Lukin, *Quantum Repeater with Encoding*, Physical Review A **79**, 032325 (2009).

[14] W. J. Munro, A. M. Stephens, S. J. Devitt, K. A. Harrison, and K. Nemoto, *Quantum Communication without the Necessity of Quantum Memories*, Nature Photonics **6**, 777 (2012).

[15] K. Nemoto, M. Trupke, S. J. Devitt, A. M. Stephens, B. Scharfenberger, K. Buczak, T. Nöbauer, M. S. Everitt, J. Schmiedmayer, and W. J. Munro, *Photonic Architecture for Scalable Quantum Information Processing in Diamond*, Physical Review X **4**, 031022 (2014).

[16] K. Azuma, K. Tamaki, and H.-K. Lo, *All-Photonic Quantum Repeaters*, Nature Communications **6**, 6787 (2015).

[17] J. Borregaard, H. Pichler, T. Schröder, M. D. Lukin, P. Lodahl, and A. S. Sørensen, *One-Way Quantum Repeater Based on Near-Deterministic Photon-Emitter Interfaces*, Physical Review X **10**, 021071 (2020).

[18] E. Togan et al., *Quantum Entanglement between an Optical Photon and a Solid-State Spin Qubit*, Nature **466**, 730 (2010).

[19] H. Bernien et al., *Heralded Entanglement between Solid-State Qubits Separated by Three Metres*, Nature **497**, 86 (2013).

[20] H. Kosaka and N. Niikura, *Entangled Absorption of a Single Photon with a Single Spin in Diamond*, Physical Review Letters **114**, 053603 (2015).

[21] S. Yang et al., *High-Fidelity Transfer and Storage of Photon States in a Single Nuclear Spin*, Nature Photonics **10**, 507 (2016).

[22] P. C. Humphreys, N. Kalb, J. P. J. Morits, R. N. Schouten, R. F. L. Vermeulen, D. J. Twitchen, M. Markham, and R. Hanson, *Deterministic Delivery of Remote Entanglement on a Quantum Network*, Nature **558**, 268 (2018).

[23] K. Tsurumoto, R. Kuroiwa, H. Kano, Y. Sekiguchi, and H. Kosaka, *Quantum Teleportation-Based State Transfer of Photon Polarization into a Carbon Spin in Diamond*, Communications Physics **2**, 74 (2019).

[24] R. Vasconcelos, S. Reisenbauer, C. Salter, G. Wachter, D. Wirtitsch, J. Schmiedmayer, P. Walther, and M. Trupke, *Scalable Spin–Photon Entanglement by Time-to-Polarization Conversion*, Npj Quantum Information **6**, 9 (2020).

[25] Y. Sekiguchi, Y. Yasui, K. Tsurumoto, Y. Koga, R. Reyes, and H. Kosaka, *Geometric Entanglement of a Photon and Spin Qubits in Diamond*, Communications Physics **4**, 264 (2021).

[26] D. L. Moehring, P. Maunz, S. Olmschenk, K. C. Younge, D. N. Matsukevich, L.-M. Duan, and C. Monroe, *Entanglement of Single-Atom Quantum Bits at a Distance*, Nature **449**, 68 (2007).

[27] S. Ritter, C. Nölleke, C. Hahn, A. Reiserer, A. Neuzner, M. Uphoff, M. Mücke, E. Figueroa, J. Bochmann, and G. Rempe, *An Elementary Quantum Network of Single Atoms in Optical Cavities*, Nature **484**, 195 (2012).

[28] J. Hofmann, M. Krug, N. Ortegel, L. Gérard, M. Weber, W. Rosenfeld, and H. Weinfurter, *Heralded Entanglement Between Widely Separated Atoms*, Science (1979) **337**, 72 (2012).

[29] A. Delteil, Z. Sun, W. Gao, E. Togan, S. Faelt, and A. Imamoğlu, *Generation of Heralded Entanglement between Distant Hole Spins*, Nature Physics **12**, 218 (2016).

[30] N. Kalb, A. Reiserer, S. Ritter, and G. Rempe, *Heralded Storage of a Photonic Quantum Bit in a Single Atom*, Physical Review Letters **114**, 220501 (2015).

[31] C. T. Nguyen et al., *Quantum Network Nodes Based on Diamond Qubits with an Efficient Nanophotonic Interface*, Physical Review Letters **123**, 183602 (2019).

[32] R. Ikuta et al., *Polarization Insensitive Frequency Conversion for an Atom-Photon Entanglement Distribution via a Telecom Network*, Nature Communications **9**, 1997 (2018).

[33] A. Tchebotareva et al., *Entanglement between a Diamond Spin Qubit and a Photonic Time-Bin Qubit at Telecom Wavelength*, Physical Review Letters **123**, 063601 (2019).

[34] T. van Leent, M. Bock, R. Garthoff, K. Redeker, W. Zhang, T. Bauer, W. Rosenfeld, C. Becher, and H. Weinfurter, *Long-Distance Distribution of Atom-Photon Entanglement at Telecom Wavelength*, Physical Review Letters **124**, 010510 (2020).

[35] K. Mannami, T. Kondo, T. Tsuno, T. Miyashita, D. Yoshida, K. Ito, K. Niizeki, I. Nakamura, F.-L. Hong, and T. Horikiri, *Coupling of a Quantum Memory and Telecommunication Wavelength Photons for High-Rate Entanglement Distribution in Quantum Repeaters*, Optics Express **29**, 41522 (2021).

[36] J. R. Maze, A. Gali, E. Togan, Y. Chu, A. Trifonov, E. Kaxiras, and M. D. Lukin, *Properties of Nitrogen-Vacancy Centers in Diamond: The Group Theoretic Approach*, New Journal of Physics **13**, 025025 (2011).

[37] N. Kalb, P. C. Humphreys, J. J. Slim, and R. Hanson, *Dephasing Mechanisms of Diamond-Based Nuclear-Spin Memories for Quantum Networks*, Physical Review A **97**, 062330 (2018).

[38] C. E. Bradley et al., *Robust Quantum-Network Memory Based on Spin Qubits in Isotopically Engineered Diamond*, arXiv [quant-ph]. (2021). Available: https://arxiv.org/abs/2111.09772